\documentclass[review,sort&compress,times]{elsarticle}

\usepackage{amsmath}
\usepackage{booktabs}
\usepackage{array}
\graphicspath{{./FIGURES/}} % figures folder

\begin{document}

\noindent NOTICE: this is the author’s version of a work that was accepted for publication in \emph{Solid-State Electronics}. Changes resulting from the publishing process, such as peer review, editing, corrections, structural formatting, and other quality control mechanisms may not be reflected in this document. Changes may have been made to this work since it was submitted for publication. A definitive version was subsequently published in \emph{Solid-State Electronics}, [VOL 100, (October 2014)] DOI 10.1016/j.sse.2014.07.003

\begin{frontmatter}

\title{Boosting the voltage gain of graphene FETs through a differential amplifier scheme with positive feedback}

\author[arces]{R. Grassi\corref{cor1}\fnref{fn1}}
\ead{rgrassi@arces.unibo.it}
\author[arces]{A. Gnudi}
\ead{agnudi@arces.unibo.it}
\author[arces]{V. Di Lecce}
\ead{vdilecce@arces.unibo.it}
\author[arces]{E. Gnani}
\ead{egnani@arces.unibo.it}
\author[arces]{S. Reggiani}
\ead{sreggiani@arces.unibo.it}
\author[arces]{G. Baccarani}
\ead{gbaccarani@arces.unibo.it}
\cortext[cor1]{Corresponding author}
\address[arces]{E. De Castro Advanced Research Center on Electronic Systems (ARCES), University of Bologna,
Viale del Risorgimento 2, 40136 Bologna, Italy}
\fntext[fn1]{Phone: +39 051 209 3772 ; Fax: +39 051 209 3779}





\begin{abstract}
We study a possible circuit solution to overcome the problem of low voltage gain of short-channel graphene FETs. The circuit consists of a fully differential amplifier with a load made of a cross-coupled transistor pair. Starting from the device characteristics obtained from self-consistent ballistic quantum transport simulations, we explore the circuit parameter space and evaluate the amplifier performance in terms of dc voltage gain and voltage gain bandwidth. We show that the dc gain can be effectively improved by the negative differential resistance provided by the cross-coupled pair. Contact resistance is the main obstacle to achieving gain bandwidth products in the terahertz range. Limitations of the proposed amplifier are identified with its poor linearity and relatively large Miller capacitance.
\end{abstract}


\end{frontmatter}

%\linenumbers

\section{Introduction}
\label{sec_intro}

Graphene has received considerable interest in recent years for radio-frequency (RF) applications \cite{Schwierz10,Schwierz13}. Its high intrinsic carrier mobility \cite{Bolotin2008} and large Fermi velocity \cite{Geim07} promise a high device transconductance $g_m$, which should in principle enable device operation up to the terahertz range of frequencies. Integrated circuits made of graphene FETs (GFETs) have already been demonstrated \cite{Wang10,Lin11} and cut-off frequencies of hundreds of gigahertz, competing with the ones of III-V HEMTs, have indeed been reported \cite{Liao10,Lin10,Wu11,Cheng12}. Despite these progresses, challenges still remain, in particular, in achieving maximum oscillation frequencies comparable with cut-off frequencies for power amplifier applications\cite{Wu12NL,Guo13} and intrinsic gains $g_m/g_d$ ($g_d$ is the device drain conductance) suitable for voltage amplifier applications \cite{Han11,Guerriero12,Wu12NL,Rizzi12}. The latter problem is due to the poor current saturation (i.e., large $g_d$) in monolayer GFETs related to the absence of a band gap \cite{Meric08,Han12}. Values of voltage gain up to 14 have actually been reported for graphene inverters made of long-channel monolayer devices (of the order of $1$~$\mu$m) \cite{Rizzi12,Schall13}. However, as the channel length is scaled down, the gain gets much worse \cite{Wu12NL} since velocity saturation due to carrier scattering, which helps current saturation, gets suppressed. 

The use of bilayer graphene has been proven to effectively improve the gain of GFETs, leading to $g_m/g_d$ values as high as 35 thanks to a band gap opening effect \cite{Szafranek12,Fiori12}. However, this kind of device inherently requires large electric fields and large voltage drops in the vertical direction. In a recent paper of ours \cite{Grassi13b} we have investigated a different approach toward the improvement of the voltage gain, which relies on the effect of negative differential resistance (NDR) occurring in the output characteristics of monolayer devices under specific bias conditions. A gain enhancement can be obtained thanks to a cancellation effect between the negative $g_d$ and the positive conductance of the load. The advantage over the bilayer GFET is that the presence of a back gate with a large applied voltage is not essential, although it can be useful in order to provide electrostatic doping of the source and drain access regions. On the other hand, the major drawback is represented by the difficulty in controlling circuit stability, which requires a proper compensation at both the input and the output ports and limits the scope of possible applications.

In the context of standard CMOS analog circuit design, a technique based on a similar concept of conductance cancellation (named ``positive-feedback'' or ``negative conductance'' technique) has been proposed to enhance the gain and gain-bandwidth-product (GBW) of amplifiers \cite{Allstot82,Amourah02,Tran12}. In this case, the output conductance is positive and a feedback from the output node is used to generate a negative load conductance that compensates its value, thereby achieving high gains. Such technique could also be applied to GFET-based amplifiers. Since the devices would operate in the standard non-NDR regime, the aforementioned stability problems would be avoided. It is the purpose of this work to investigate such an idea, through a numerical study of the small-signal performance of a specific positive-feedback amplifier made by short-channel GFETs.

The paper is organized as follows. Section~\ref{sec_device} deals with the reference GFET that is used as building block of the amplifier: the device structure, simulation model, and device characteristics are briefly discussed. In Section~\ref{sec_amplifier} the amplifier circuit is presented. Therein, a detailed circuit analysis, based on the simulated characteristics of the single device and focused on the amplifier dc gain and gain bandwidth, is reported. Conclusions are finally drawn in Section~\ref{sec_conclusions}.

\section{Structure, model, and simulation results of the intrinsic device}
\label{sec_device}

\begin{figure}[!t]
\centering
\includegraphics[scale=0.63]{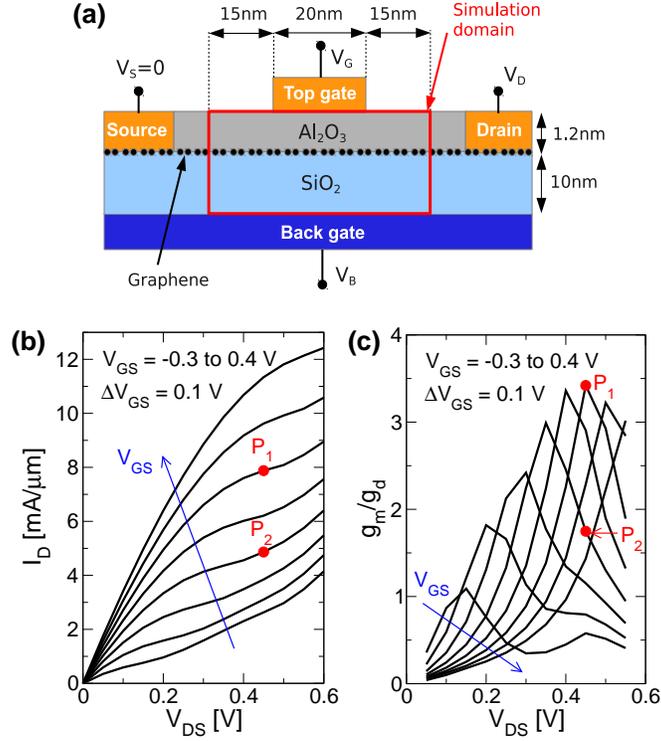}
\caption{(a) Longitudinal cross-section of the simulated GFET. (b) Output characteristics at $V_{BS}=9$~V showing n-type-FET operation. (c) Corresponding intrinsic voltage gain. A zero top-gate/graphene workfunction difference $\Phi_{mg}$ is assumed. $P_1$ and $P_2$ are specific operating points. Figs. (a--b) are adapted from \cite{Grassi13b}.}
\label{fig_intrinsic_device}
\end{figure}
The device is identical to the one considered in our previous work \cite{Grassi13b}. The structure is a dual-gate GFET, with a top gate length of $20$~nm (Fig.~\ref{fig_intrinsic_device}a). The top dieletric is a 1.2-nm-thick layer of Al$_2$O$_3$ (effective oxide thickness of 0.5~nm), whereas the back dielectric is silicon oxide with thickness of $10$~nm. The back gate is used as a means to electrostatically dope the graphene access regions between the top gate and the source and drain contacts, and it could be replaced by a second top gate covering the access regions alone.

The dc characteristics of the device are computed with an in-house developed code for GFETs, based on the self-con\-sis\-tent solution of the 2-D Poisson equation and the ballistic non-e\-qui\-lib\-ri\-um Green's function (NEGF) equations \cite{DattaQT2005}, with a $p_z$ tight-binding Hamiltonian (more details can be found in \cite{Grassi13b}).

The small-signal frequency analysis is based on a small-signal equivalent circuit (see \cite[Fig.~2]{Grassi13b}) whose intrinsic part is derived from \cite{Tsividis99}. The intrinsic small-signal circuit parameters are extracted from the simulated dc $I$--$V$ and $Q$--$V$ characteristics using finite differences. As for the parasitics, we consider the possibility of non-zero source and drain contact resistances; sensitivity to gate resistance and parasitic capacitances is considered for one specific case.

Fig.~\ref{fig_intrinsic_device}b shows the device output characteristics obtained with a back-gate-to-source voltage $V_{BS}=9$~V and assuming a zero metal-graphene workfunction difference for both gates (in this paper the back-gate/graphene workfunction difference will always be taken to be zero). Such value of $V_{BS}$ corresponds to a heavy n-type doping of the source and drain regions. In all the range of top-gate-to-source voltage $V_{GS}$ considered in the plot, the channel doping is also n-type and the device operates as a conventional n-type FET, although showing only a weak (or ``quasi-'') saturation. The resulting intrinsic gain is limited to about 3.4 (Fig.~\ref{fig_intrinsic_device}c). Thanks to the symmetry of the graphene bandstructure and the use of an electrostatic doping, the same device exhibits p-type characteristics completely symmetrical to the ones in Fig.~\ref{fig_intrinsic_device}b if the polarities of all voltages, including $V_{BS}$, are reversed (not shown).

\section{Analysis of the ``positive-feedback'' amplifier}
\label{sec_amplifier}

\begin{figure}[!t]
\begin{minipage}[c]{0.55\linewidth}
\centering
\includegraphics[scale=0.63]{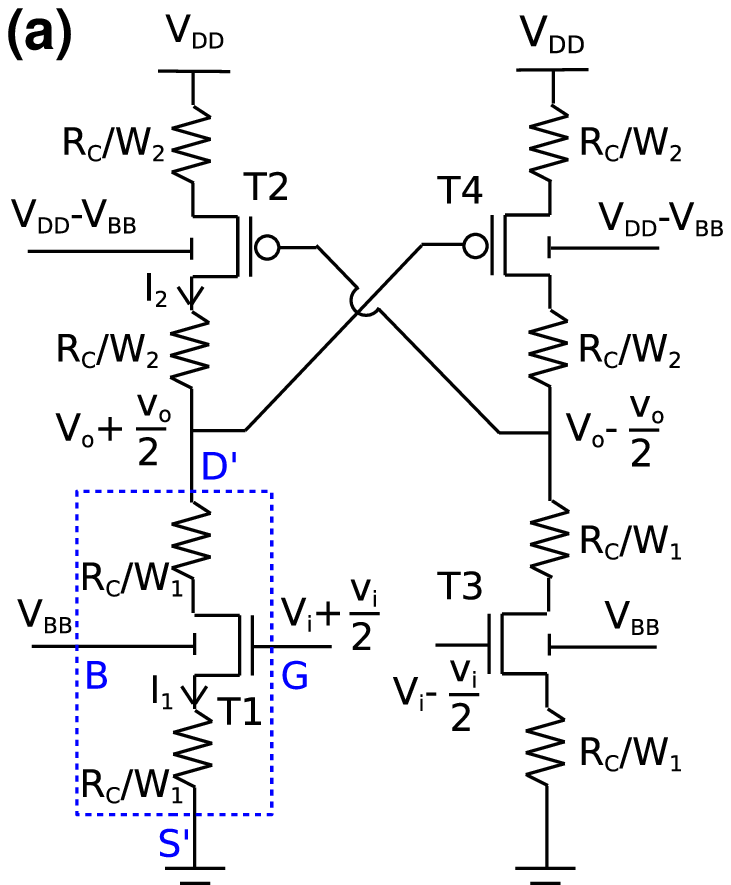}
\end{minipage}
\begin{minipage}[c]{0.44\linewidth}
\centering
\includegraphics[scale=0.63]{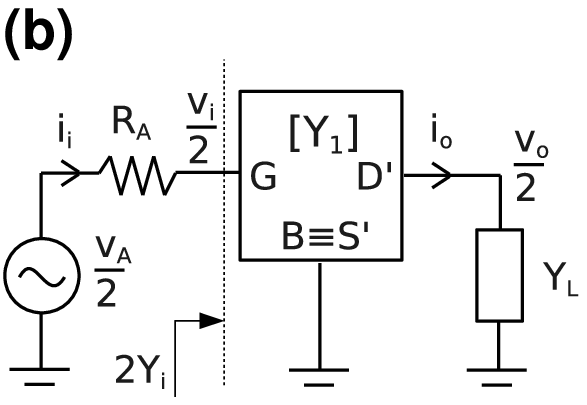}
\small
\begin{multline*}
Y_L=[Y_2]_{11}+[Y_2]_{22}+{}\\
{}-[Y_2]_{12}-[Y_2]_{21}
\end{multline*}
\end{minipage}
\caption{(a) Schematic of the CS differential amplifier. Transistors T1 and T2 are identical to T3 and T4, respectively. Resistors represent source and drain contact resistances. (b) Corresponding small-signal model, where $[Y_1]$ ($[Y_2]$) is the $Y$-matrix of the extrinsic transistor T1 (T2). $Y_L$ is calculated from $[Y_2]$ as indicated. $R_A$ is the output resistance of the input voltage source.}
\label{fig_circuit}
\end{figure}
The amplifier circuit considered here (Fig.~\ref{fig_circuit}a) is similar to the first one proposed in the literature on positive-feedback amplifiers (see \cite[Fig.~1a]{Amourah02}). It is a common-source (CS) fully differential amplifier with n-type driver transistors and an active load made of cross-coupled p-type transistors. Although different symbols are used in Fig.~\ref{fig_circuit}a to indicate n- and p-type transistors, they actually share the same structure: as explained in Section~\ref{sec_device}, one or the other type of transistor is obtained by changing the polarity of $V_{BS}$. The circuit is symmetric, in the sense that transistors T1 and T2 are respectively identical to T3 and T4. On the other hand, T1 and T2 may have different channel widths ($W_1$ and $W_2$, respectively) or different top gate materials, resulting in different top-gate/graphene workfunction differences ($\Phi_{mg1}$ and $\Phi_{mg2}$, respectively, with $\Phi_{mg1}$ assumed zero). Contact resistances are treated as fixed parameters independent of bias. We consider a common value for the source and drain contact resistances of the same transistor (because of the symmetry of the device structure), and we assume it to scale with the inverse of the channel width ($1/R_C$ is the contact conductance per unit width). Unprimed/primed symbols are used to indicate intrinsic/extrinsic device terminals, as done in Fig.~\ref{fig_circuit}a for T1 (the blue box represents the ``extrinsic'' device T1). The circuit equations are solved using a spline interpolation of the device $I$--$V$ characteristics obtained in Section~\ref{sec_device}.

Thanks to the symmetry of the amplifier, a simple small-signal equivalent circuit is derived for the case of purely differential (zero common mode) signals (Fig.~\ref{fig_circuit}b), which involves only the $Y$-matrices of transistors T1 and T2 ($[Y_1]$ and $[Y_2]$, respectively), which, in turn, are easily calculated from the device small-signal equivalent circuit.

According to the symbol definitions in Fig.~\ref{fig_circuit}b, the amplifier voltage gain is given by $A_v=v_o/v_i=-[Y_1]_{21}/([Y_1]_{22}+Y_L)$ and takes a simple form at frequency $f=0$:
\begin{equation} \label{eq_av}
A_{v0} = - \frac{\tilde{g}_{m1}}{\tilde{g}_{d1}+\tilde{g}_{d2}-\tilde{g}_{m2}} ,
\end{equation}
where $g_{m1}/\tilde{g}_{m1}=g_{d1}/\tilde{g}_{d1}=[1+(R_C/W_1)(2 g_{d1}+g_{m1}+g_{mb1})]$, with $g_{d1}$, $g_{m1}$, and $g_{mb1}$ the drain conductance, top-gate transconductance, and back-gate transconductance of T1, respectively (similar definitions hold for T2). Eq.~\ref{eq_av} explains the gain-en\-hance\-ment effect: the term $G_L=\tilde{g}_{d2}-\tilde{g}_{m2}$, which is the load conductance due to T2, can be negative and compensate $\tilde{g}_{d1}$. The gain is theoretically infinite if the matching is exact. However, the circuit becomes unstable for $\tilde{g}_{d1}+G_L<0$ (or $A_{v0} > 0$) \footnote{We are referring here to the stability of the amplifier connected to a short circuit at the input port and to an open circuit at the output port. The stability in other source/load conditions can be evaluated with the standard stability-circle technique \cite{Pozar05}.}. If conventional well-saturated MOSFETs were used, $g_m \gg g_d$, hence additional diode-connected transistors in parallel to the cross-coupled load transistors would be required to allow circuit stability \cite[Fig.~1a]{Amourah02}. These are not necessary in our GFET implementation since $g_m \sim g_d$ (Fig.~\ref{fig_intrinsic_device}c).

\subsection{Dc voltage gain and voltage gain bandwidth}

The circuit operating point is determined by a considerable numbers of variables/pa\-ram\-e\-ters, e.g., $V_i$, $V_{BB}$, $V_{DD}$, $W_2/W_1$, $\Phi_{mg2}$ and $R_C$ (see Fig.~\ref{fig_circuit}a; $W_1$ can be taken as the reference width). In order to simplify the analysis, we choose to fix the operating point of T1, for example, at the point $P_1$ of maximum $g_m/g_d$ of Figs.~\ref{fig_intrinsic_device}b--c, which, according to (\ref{eq_av}), should favor high amplifier gains since it also corresponds to high $g_m$ ($\simeq 17$ mS/$\mu$m). In this way, for a given value of $R_C$, we fix $V_i$, $V_{BB}$, as well as the output voltage $V_o$ and the current $I_1$ that flows through T1. For a given value of $\Phi_{mg2}$, a relation between $V_{DD}$ (or $V_L=V_{DD}-V_o$) and $W_2$ is obtained from the $I$--$V$ characteristics of T2 and the constraint $I_2 = I_1$.

\begin{figure}[!t]
\centering
\includegraphics[scale=0.66]{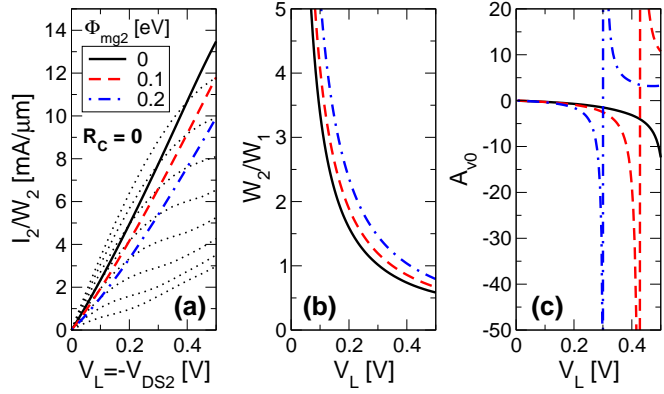}
\caption{Analysis of the dc voltage gain as a function of $V_L=V_{DD}-V_o$. The operating point of T1 is held constant at $P_1$ and a zero contact resistance is assumed. (a) $I$--$V$ characteristics of diode-connected T2 for different values of top-gate/graphene workfunction difference $\Phi_{mg2}$. The dotted lines show the T2 output characteristics at constant $V_{GS2}+\Phi_{mg2}$ from -0.4~V (top curve) to 0.3~V (bottom curve) in steps of 0.1~V. (b) Corresponding ratio between the widths of T1 and T2, which is determined by the constraint $I_1=I_2$. (c) Corresponding dc voltage gain. Positive values indicate an unstable circuit.}
\label{fig_I2_Wratio_Av0_vs_vl}
\end{figure}
The above procedure is illustrated in Fig.~\ref{fig_I2_Wratio_Av0_vs_vl}, where we take the bias point of T1 equal to $P_1$, $R_C=0$, and we consider three values of $\Phi_{mg2}$. For each value of $V_{DD}$, we first compute the current per unit width through T2, $I_2/W_2$, by imposing its top-gate and drain voltages to be equal (Fig.~\ref{fig_I2_Wratio_Av0_vs_vl}a). Then, we calculate $W_2/W_1$ from $W_2/W_1=(I_1/W_1)/(I_2/W_2)$ (Fig.~\ref{fig_I2_Wratio_Av0_vs_vl}b). Finally, having set the operating points of both T1 and T2, we obtain the amplifier gain $A_{v0}$ using (\ref{eq_av}) (Fig.~\ref{fig_I2_Wratio_Av0_vs_vl}c). It can be observed that, if $\Phi_{mg2}=0$ and $V_L$ is limited to 0.5~V, the maximum gain is only 12, which means that $G_L$ is not sufficiently negative. In order to further decrease $G_L$, which is reduced to $G_L=g_{d2}-g_{m2}$ in the case of $R_C=0$, one needs to increase the ratio $g_{m2}/g_{d2}$, i.e., to move the operating point of T2 into the quasi-saturation region. This can be achieved by choosing a suitable top-gate material of T2 that provides $\Phi_{mg2}>0$, as shown in Fig.~\ref{fig_I2_Wratio_Av0_vs_vl}a, where the T2 output characteristics at constant $V_{GS2}+\Phi_{mg2}$ are also shown for comparison. The considered values of $\Phi_{mg2}$ are within the range of workfunction of several metals \cite{Haynes13}. Increasing $\Phi_{mg2}$ also leads to the aforementioned divergence and change of sign of $A_{v0}$, which mark the beginning of the instability region (Fig.~\ref{fig_I2_Wratio_Av0_vs_vl}c). It should be noted that the resulting values of $W_2/W_1$ are reasonable in all three cases.

\begin{figure}[!t]
\centering
\includegraphics[scale=0.66]{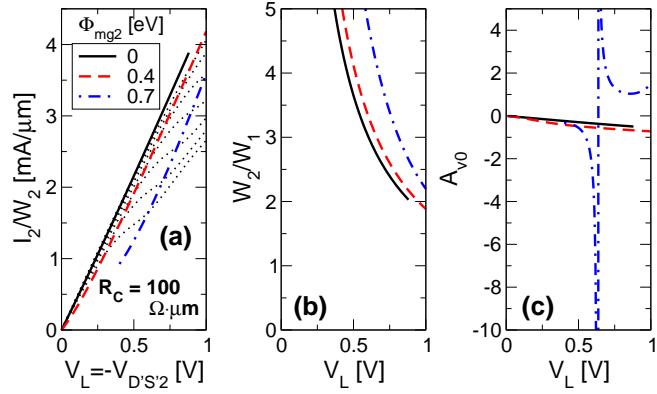}
\caption{Same as in Fig.~\ref{fig_I2_Wratio_Av0_vs_vl} but for $R_C=100$~$\Omega\cdot\mu$m. Here, $V_{BB}=9.79$~V. The dotted lines in (a) are the output characteristics of the extrinsic transistor T2 at constant $V_{GS'2}+\Phi_{mg2}$ from -0.5~V (top curve) to 0.1~V (bottom curve) in steps of 0.1~V.}
\label{fig_I2_Wratio_Av0_vs_vl_Gs_Gd}
\end{figure}
The same procedure has been repeated for the case of $R_C=100$~$\Omega\cdot\mu$m (Fig.~\ref{fig_I2_Wratio_Av0_vs_vl_Gs_Gd}), a value which is within reach of present graphene technology \cite{Moon12}. It is seen that, given a maximum $V_L$ voltage of 1~V, it is still possible to achieve high gains, but a larger $\Phi_{mg2}$ is required to move the bias point of T2 into the quasi-saturation region ($\Phi_{mg2}\simeq 0.7$~eV could be achieved with, for example, Au, Pd, or Pt \cite{Haynes13}).

The amplifier bandwidth is evaluated by numerically computing the frequency $f_p$ of the dominant pole of $A_v(s)=v_o(s)/v_i(s)$. In the case of $R_C=0$, an analytical expression is available:
\begin{equation} \label{eq_fp}
f_p=\frac{1}{2\pi}\frac{g_{d1}+g_{d2}-g_{m2}}{C_{dd1}+C_L} ,
\end{equation}
where $C_L=C_{gg2}+C_{dd2}+C_{gd2}+C_{dg2}$ (the subscripts ``1'' and ``2'' refer respectively to T1 and T2 as usual), and the capacitances are defined as $C_{kk}=\partial Q_{K}/\partial V_{K}$ and $C_{kl}|_{k\ne l}=-\partial Q_{K}/\partial V_{L}$ ($k,l\in \{g,d,s,b\}$), with the charges $Q_K$ computed as described in \cite{Grassi13b}. The gain-bandwidth-product is defined as $\text{GBW}=|A_{v0}|f_p$.

\begin{figure}[!t]
\centering
\includegraphics[scale=0.65]{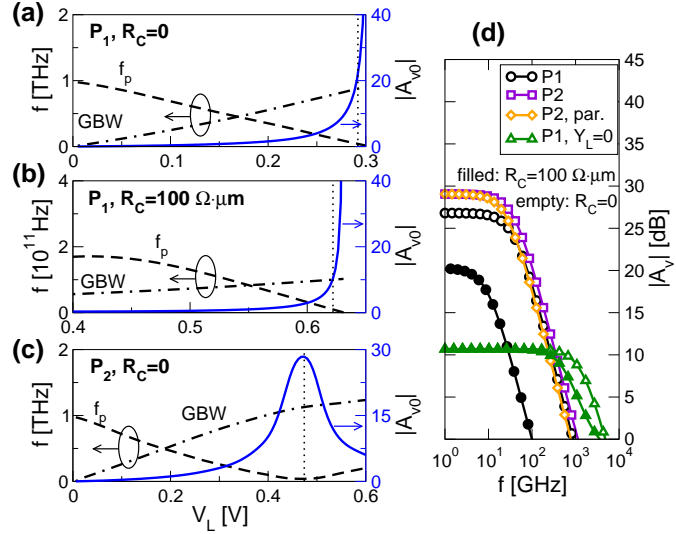}
\caption{(a--c) Dc voltage gain, dominant pole frequency, and gain bandwidth product vs. $V_L$ for different operating points of T1 and values of contact resistance (see legend). Values of $\Phi_{mg2}$ in (a--c) are 0.2, 0.7, and 0.2~eV, respectively. (d) Frequency magnitude response of voltage gain at the $V_L$ points indicated by vertical dotted lines in (a--c). The performance of transistor T1 with open circuit load ($Y_L=0$) is also shown for comparison. The curve labeled ``P2, par.'' is obtained at the operating point $P_2$ including gate resistance and parasitic capacitances as described in the text.}
\label{fig_Av0_fp_GBW}
\end{figure}
The plots of $f_p$ and GBW for the cases of $R_C=0$ and $\Phi_{mg2}=0.2$~eV and of $R_C=100$~$\Omega\cdot\mu$m and $\Phi_{mg2}=0.7$~eV are shown in Figs.~\ref{fig_Av0_fp_GBW}a--b, respectively, where we also replot the respective dc gain from Figs.~\ref{fig_I2_Wratio_Av0_vs_vl}--\ref{fig_I2_Wratio_Av0_vs_vl_Gs_Gd}. Contact resistance has a dramatic effect on both $f_p$ and GBW, reducing them by almost one order of magnitude.

It is interesting to see how the RF performance is affected by choosing a different bias point for T1, for example, the point $P_2$ of Figs.~\ref{fig_intrinsic_device}b--c, which corresponds to a lower $g_m$ ($\simeq 11.9$~mS$/\mu$m). As shown in Fig.~\ref{fig_Av0_fp_GBW}c for the case of $R_C=0$, the performance is actually improved. Higher values of GBW (exceeding 1~THz) can be achieved, which means that the lower $g_m$ gets compensated by smaller capacitances. In addition, the dc gain is bounded, which indicates that the circuit is stable over the whole considered range of $V_L$.

The frequency magnitude response of $A_{v}$ for the three cases in Figs.~\ref{fig_Av0_fp_GBW}a--c and at specific $V_L$ points is shown in Fig.~\ref{fig_Av0_fp_GBW}d. There, we also include the voltage gain that one would obtain without recurring to positive feedback, i.e., of transistor T1 with open circuit load ($Y_L\equiv0$). Moreover, we include the results obtained by repeating one of the simulations with non-zero gate resistance $R_g$ and non-zero parasitic capacitances $C_{int}$ and $C_{ext}$ between intrinsic and extrinsic gate and source and drain terminals, respectively. The values of the parasitics of T1 have been taken equal to the ones in \cite{Grassi13b} ($R_g=4$~$\Omega$ and $C_{int}=C_{ext}=0.1$~fF), which were estimated for a channel width $W=1$~$\mu$m. The values of the parasitics of T2 have been fixed accordingly, assuming the scaling relations $R_g, C_{int}, C_{ext} \propto W$.

\begin{table}[!t]
\footnotesize
\renewcommand{\arraystretch}{1.3}
\caption{RF metrics from Fig.~\ref{fig_Av0_fp_GBW}d.}
\label{tab_rf}
\centering
\begin{tabular}{c c c >{\centering}m{0.5cm} >{\centering}m{1cm} c c}
\hline
& \multicolumn{4}{c}{$R_C=0$ $\Omega\cdot\mu$m} & \multicolumn{2}{c}{$R_C=100$ $\Omega\cdot\mu$m} \\
\cmidrule(lr){2-5}\cmidrule(lr){6-7}
& $P_1$ & $P_2$ & $P_2$, par. & $P_1$, $Y_L=0$ & $P_1$ & $P_1$, $Y_L=0$\\ 
\hline
$|A_{v0}|$ & 21.9 & 28.4 & 28.4 & 3.42 & 10.3 & 3.42\\
$f_p$ [GHz] & 40.1 & 39.8 & 27.4 & 1460 & 9.7 & 585\\
GBW [GHz] & 878 & 1130 & 777 & 4990 & 100 & 2000\\
\hline
\end{tabular}
\end{table}
The figures of merit extracted in the different cases are reported in Table~\ref{tab_rf}. It is seen that the load admittance introduced by the positive feedback significantly degrades GBW and also increases the sensitivity of GBW to contact resistance.  
%It is seen that, if $Y_L$ were zero, GBW would be in the terahertz range even with $R_C=100$~$\Omega\cdot\mu$m. Hence, the strong sensitivity of GBW to contact resistance is peculiar to the positive-feedback circuit.
The effect of gate resistance and parasitic capacitances is not as drastic as contact resistance since the corresponding degradation of GBW is only $31 \%$.

It is worth noting that, in all three cases considered in Figs. \ref{fig_Av0_fp_GBW}a--c, the qualitative trends of $f_p$ and GBW with respect to $V_L$ are similar: $f_p$ decreases with increasing $|A_{v0}|$, whereas GBW monotonically increases with $V_L$. This can be understood, at least in the case of $R_C=0$, with the help of (\ref{eq_av}) and (\ref{eq_fp}). Combining them, one obtains $f_p \propto |A_{v0}|^{-1} (C_{dd1}+C_L)^{-1}$, which implies $\text{GBW}\propto(C_{dd1}+C_L)^{-1}$. According to the latter expression, since GBW is found to increase with $V_L$, the capacitance $C_L$ related to T2 must correspondingly decrease.
\begin{figure}[!t]
\centering
\includegraphics[scale=0.6]{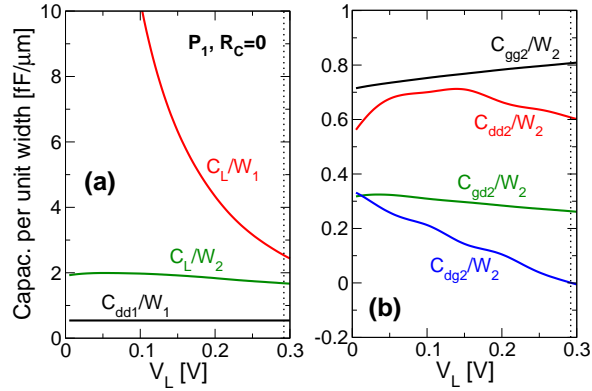}
\caption{Small-signal capacitances for the case considered in Fig.~\ref{fig_Av0_fp_GBW}a. (a) Dependence on $V_L$ of $C_{dd1}$ and $C_L$, which contribute to $f_p$ according to Eq.~\ref{eq_fp}. (b) Dependence on $V_L$ of the different contributions to $C_L$. The vertical dotted lines are located as in Fig.~\ref{fig_Av0_fp_GBW}a.}
\label{fig_capacitance}
\end{figure}
This is confirmed by the analysis of the various capacitances in Fig.~\ref{fig_capacitance}, where we have considered the case of bias point $P_1$ and $R_C=0$. The change in $C_L/W_1$ with respect to $V_L$ is caused by both the change of operating point of T2 and the adjustment of the ratio $W_2/W_1$ (recall Figs.~\ref{fig_I2_Wratio_Av0_vs_vl}a--b). Comparing $C_L/W_1$ with $C_L/W_2$, which, on the contrary, is independent of $W_2/W_1$, it is clear that the latter cause is predominant. 

\subsection{Low-voltage operation}

The strategy followed so far to find the optimal bias point of the amplifier, which consists in keeping the operating point of T1 fixed and moving $V_{L}$, uses $V_{DD}$ as a freely adjustable parameter. For example, a value of $V_{DD}\simeq0.92$~V is required to bias the amplifier in the point of maximum gain of Fig.~\ref{fig_Av0_fp_GBW}c. In order to verify if the proposed circuit can work at lower supply voltage, we have solved the circuit equations for a fixed $V_{DD}=0.55$~V, sweeping the input voltage $V_i$ for different $W_2/W_1$ ratios.
\begin{figure}[!t]
\centering
\includegraphics[scale=0.63]{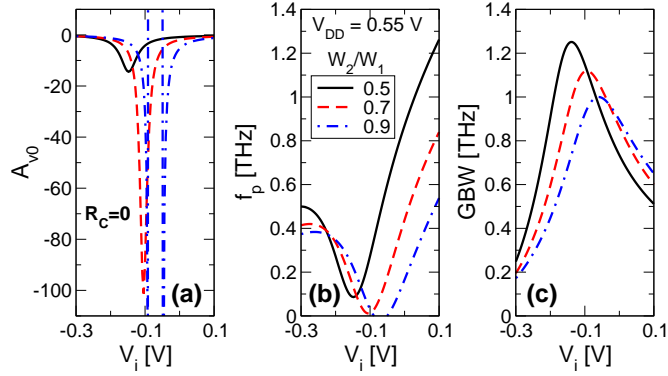}
\caption{Dc voltage gain (a), dominant pole frequency (b), and gain bandwidth product (c) as a function of $V_i$. Here the supply voltage is held constant at 0.55~V and three different values of $W_2/W_1$ are considered. A zero contact resistance is assumed. The other parameters are: $V_{BB}=9$~V and $\Phi_{mg2}=0.3$~V.}
\label{fig_Av0_vs_vi}
\end{figure}
The results are shown in Fig.~\ref{fig_Av0_vs_vi}. It is seen that, at least if $R_C=0$, it is still possible to find a bias point which provides at the same time both high gain ($>10$) and GBW above 1~THz.

\subsection{Non-linearity}

Let $V_{i1}$ and $V_{i2}$ be the voltages applied to the gates of T1 and T3, respectively, which are reduced to $V_i+v_i/2$ and $V_i-v_i/2$ in the case of a small purely differential signal (Fig.~\ref{fig_circuit}a). The strong dependence of $A_{v0}$ on $V_L$ and $V_i$ shown in Figs.~\ref{fig_I2_Wratio_Av0_vs_vl}--\ref{fig_Av0_fp_GBW} and \ref{fig_Av0_vs_vi} indicates a strong non-linearity. We have verified this by computing the differential voltage transfer characteristics for the case in Fig.~\ref{fig_Av0_vs_vi} with $W_2/W_1=0.7$ and a value of $V_i=-0.1$~V close to the point of maximum gain.  
\begin{figure}[!t]
\centering
\includegraphics[scale=0.28]{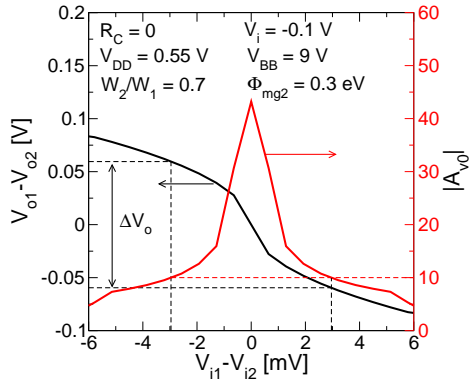}
\caption{Differential voltage transfer characteristics and corresponding voltage gain obtained with the parameters indicated in the legend. The common-mode input voltage is chosen close to the point of maximum gain of the curve with $W_2/W_1=0.7$ in Fig.~\ref{fig_Av0_vs_vi}a. $\Delta V_o$ is defined in the text.}
\label{fig_voutd_vind}
\end{figure}
As shown in Fig.~\ref{fig_voutd_vind}, the strong sensitivity of the gain to the input voltage results in a small high-gain output voltage range $\Delta V_o$ compared to $V_{DD}$: for example, defining $\Delta V_o$ as the output voltage range where $|A_{v0}|>10$, we have $\Delta V_o/V_{DD}\simeq 0.22$. The problem is inherently related to the compensation technique and it was also found to affect NDR GFETs \cite{Grassi13b}. In the literature on positive-feedback amplifiers, some more complex versions of the amplifier have been proposed to reduce non-linearity \cite{Amourah02}. 

\subsection{Miller effect}

Thus far, we have not taken into account the output resistance $R_A$ of the input voltage source (see circuit of Fig.~\ref{fig_circuit}b). If $R_A$ is not sufficiently small, the bandwidth of the total voltage gain $v_o/v_A$ can be limited by the bandwidth of $v_i/v_A=(1+2 R_A Y_i)^{-1}$, which depends on the amplifier input admittance $Y_i$ (the symbols are defined in Fig.~\ref{fig_circuit}b). In the case of $R_C=0$, it can be shown that, at low frequency, $Y_i$ takes the form:
\begin{equation} \label{eq_yi}
Y_i(s)\sim s C_i,\quad C_i=(C_{gg1}+C_{gd1}|A_{v0}|)/2 .
\end{equation}
Contrary to traditional MOSFETs, where $C_{gd}$ is only due to conventional electrostatic drain-induced barrier lowering, a pronounced drain quantum capacitance arising from the lack of a band gap also contributes to $C_{gd}$ in GFETs \cite{Holland13}. For example, $C_{gd1}/C_{gg1}\simeq 0.39$ at the operating point $P_2$ (the values per unit width of the two capacitances are similar to those shown in Fig.~\ref{fig_capacitance}b for T2). The expression in (\ref{eq_yi}), although valid only at low frequency, suggests that a high dc gain $|A_{v0}|$ might have a negative impact on the bandwidth of $v_o/v_A$ through a large $Y_i$. This is the well-known Miller effect, which is typical of CS amplifiers.
\begin{figure}[!t]
\centering
\includegraphics[scale=0.63]{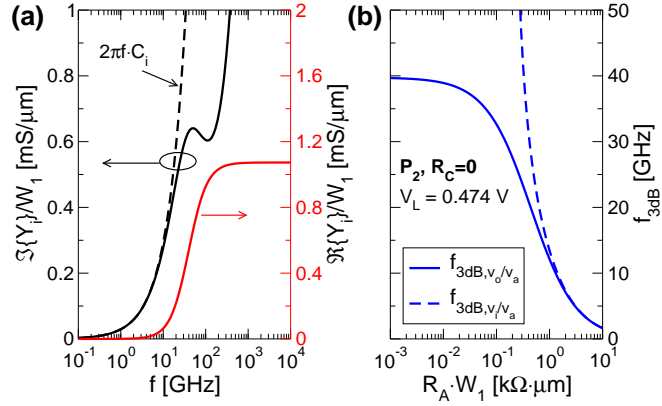}
\caption{(a) Real and imaginary parts of the input admittance $Y_i$ as a function of frequency (solid lines) for the case considered in Fig.~\ref{fig_Av0_fp_GBW}c with $V_L=0.474$~V. The dashed line is the plot of $2 \pi f C_i$ with $C_i=4.7$~fF/$\mu$m. (b) $-$3dB frequency of voltage gains $v_o/v_A$ and $v_i/v_A$ as a function of the source resistance $R_A$ (see Fig.~\ref{fig_circuit}b for the definitions).}
\label{fig_miller}
\end{figure}
The real and imaginary parts of the normalized $Y_i$ are shown in Fig.~\ref{fig_miller}a for the case in Fig.~\ref{fig_Av0_fp_GBW}c with $V_L=0.474$~V. It is seen that the approximation in (\ref{eq_yi}) represents a good fit of $Y_i$ up to about 10~GHz. To evaluate the impact of $\Im{\{Y_i\}}$ on the amplifier bandwidth, we numerically compute the $-$3dB frequency of $v_o/v_A$ as a function of $R_A$ (Fig.~\ref{fig_miller}b). The bandwidth drops already at $R_A W_1 \sim 100$~$\Omega\cdot \mu$m, indicating that the input susceptance is relatively large.

\subsection{Common-gate variant}

\begin{figure}[!t]
\centering
\includegraphics[scale=0.63]{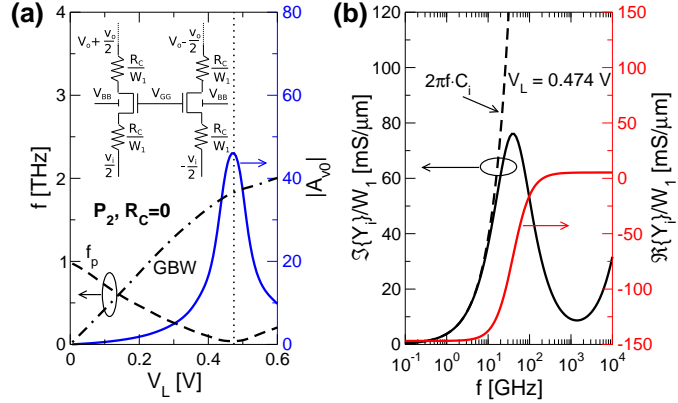}
\caption{(a) Same as in Fig.~\ref{fig_Av0_fp_GBW}c but for the CG configuration (the input transistor connection is shown in the inset). The RF metrics at $V_L=0.474$~V are: $|A_{v0}|=46.1$, $f_p=39.8$~GHz, and $\textrm{GBW}=1.83$~THz. (b) Same as in Fig.~\ref{fig_miller}a but for the CG configuration. Here, $C_i=610$~fF/$\mu$m.}
\label{fig_cg}
\end{figure}
Since Miller's effect is usually mitigated in the common-gate (CG) configuration, we have considered a CG version of the amplifier (the input transistor connection is shown in the inset of Fig.~\ref{fig_cg}a, whereas the cross-coupled load transistors are identical to the CS case) and repeated the analysis of the RF performance with varying $V_L$ at a fixed operating point of T1, using the same parameters as in Fig.~\ref{fig_Av0_fp_GBW}c. The results in Fig.~\ref{fig_cg}a show that the CG amplifier has identical $f_p$ and slightly higher $A_{v0}$ and GBW compared to the CS amplifier. Indeed, in the CG configuration and in the case of $R_C=0$, the dc gain is given by $A_{v0}=(g_{m1}+g_{mb1}+g_{d1})/(g_{d1}+g_{d2}-g_{m2})$, which differs from the CS expression because of the additional term $g_{mb1}+g_{d1}\sim g_{d1}$ in the numerator. Unfortunately, the CG amplifier is not unilateral at low frequency due to the high value of $g_d$ between the input and output ports. Looking at the plot of the real part of $Y_i$ in Fig.~\ref{fig_cg}b, it is clear that the negative load conductance gives rise to a negative input conductance. Indeed, it can be shown that $Y_i|_{f=0}=A_{v0}G_L$. Because of the negative $\Re\{Y_i\}$ at low frequency, the circuit is unstable for $R_A W_1>6.8$~$\Omega\cdot\mu$m. Notwithstanding the $\Im\{Y_i\}$ values, the CG configuration is thus ruled out.

\section{Conclusions}
\label{sec_conclusions}

In this work, we have investigated the use of the positive-feedback technique to increase the voltage gain of GFET-based amplifiers in CS configuration. The RF performance of the reference amplifier has been evaluated through a small-signal analysis with parameters extracted from atomistic quantum transport simulations. The analysis has shown that, with a proper choice of the relative widths of the transistors and of the gate metal workfunctions, it is possible to find bias points with high dc gain ($>10$), even at relatively small supply voltages. On the other hand, the dc gain is strongly sensitive to the bias point, a problem which is common to other conductance compensation techniques. Contact resistance at typical experimental values has been found to significantly degrade GBW, which would be otherwise in the terahertz range. The amplifier bandwidth can also be negatively affected by Miller's effect, if the output resistance of the input voltage source is not sufficiently small. The use of the CG configuration to circumvent the Miller effect is ruled out due to stability problems.

\section*{Acknowledgments}

This work has been supported by the EU project GRADE 317839.

\section*{References}

\bibliography{mybibfile}

\end{document}